\documentclass[conference]{IEEEtran}
\usepackage{cite}
\usepackage{amsthm}
\usepackage{amsmath,amssymb,amsfonts}
\usepackage{geometry}
\usepackage{graphicx}
\usepackage{textcomp}
\usepackage{xcolor}
\usepackage{bm}
\usepackage{graphicx,epstopdf,psfrag,url,cite,xcolor,multirow,float}
\usepackage[font=footnotesize]{subfig}
\usepackage{amsmath}
\usepackage{bbm}
\usepackage{algorithm}
\usepackage{algorithmic}
\usepackage{array}
\usepackage{booktabs}
\usepackage{times}
\begin{document}
%
% paper title
% Titles are generally capitalized except for words such as a, an, and, as,
% at, but, by, for, in, nor, of, on, or, the, to and up, which are usually
% not capitalized unless they are the first or last word of the title.
% Linebreaks \\ can be used within to get better formatting as desired.
% Do not put math or special symbols in the title.

% \title{Unified Codec Framework for Joint Source-Channel Orthogonality Control in IRS assisted Multi-Uuser Semantic Communications}
\title{Learning Joint Source-Channel Encoding in IRS-assisted Multi-User Semantic Communications}

%
%
% author names and IEEE memberships
% note positions of commas and nonbreaking spaces ( ~ ) LaTeX will not break
% a structure at a ~ so this keeps an author's name from being broken across
% two lines.
% use \thanks{} to gain access to the first footnote area
% a separate \thanks must be used for each paragraph as LaTeX2e's \thanks
% was not built to handle multiple paragraphs
%
\vspace{-2cm}

\author{Haidong Wang\IEEEauthorrefmark{1}\IEEEauthorrefmark{2}, Songhan Zhao\IEEEauthorrefmark{1}\IEEEauthorrefmark{2}, Lanhua Li\IEEEauthorrefmark{1}\IEEEauthorrefmark{2}, Bo Gu\IEEEauthorrefmark{1}\IEEEauthorrefmark{2}, Jing Xu\IEEEauthorrefmark{3}, Shimin Gong\IEEEauthorrefmark{1}\IEEEauthorrefmark{2}, and Jiawen Kang\IEEEauthorrefmark{4}\\
%\IEEEauthorblockA{
\IEEEauthorrefmark{1}School of Intelligent Systems Engineering, Shenzhen Campus of Sun Yat-sen University, China\\
\IEEEauthorrefmark{2}Guangdong Provincial Key Laboratory of Fire Science and Intelligent Emergency Technology, China\\
\IEEEauthorrefmark{3}School of Electronic Information and Communications, Huazhong University of Science and Technology, China\\
\IEEEauthorrefmark{4}School of Automation, Guangdong University of Technology, China\\
%%\IEEEauthorrefmark{5}School of Computer Science and Engineering, Nanyang Technological University, Singapore
%}
%\thanks{Songhan Zhao, Yusi Long, Bo Gu, and Shimin Gong are with the School of Intelligent Systems Engineering, Sun Yat-sen University, China (e-mails: \{zhaosh55, longys\}@mail2.sysu.edu.cn, \{gubo,gongshm5\}@mail.sysu.edu.cn). Nguyen Cong Luong is with the Faculty of Computer Science, Phenikaa University, Vietnam (e-mail: luong.nguyencong@phenikaa-uni.edu.vn). Bin Lyu is with the School of Communications and Information Engineering, Nanjing University of Posts and Telecommunications, China (e-mail: blyu@njupt.edu.cn).}
\vspace{-1cm}
}

% make the title area
\maketitle
% As a general rule, do not put math, special symbols or citations
% in the abstract or keywords.
\begin{abstract}
% semantic orthogonal multiple access (JSCE)
In this paper, we investigate a joint source-channel encoding (JSCE) scheme in an intelligent reflecting surface (IRS)-assisted multi-user semantic communication system. Semantic encoding not only compresses redundant information, but also enhances information orthogonality in a semantic feature space. Meanwhile, the IRS can adjust the spatial orthogonality, enabling concurrent multi-user semantic communication in densely deployed wireless networks to improve spectrum efficiency. 
We aim to maximize the users' semantic throughput by jointly optimizing the users' scheduling, the IRS's passive beamforming, and the semantic encoding strategies. To tackle this non-convex problem, we propose an explainable deep neural network-driven deep reinforcement learning (XD-DRL) framework. Specifically, we employ a deep neural network (DNN) to serve as a joint source-channel semantic encoder, enabling transmitters to extract semantic features from raw images. By leveraging structural similarity, we assign some DNN weight coefficients as the IRS's phase shifts, allowing simultaneous optimization of IRS's passive beamforming and DNN training. Given the IRS's passive beamforming and semantic encoding strategies, user scheduling is optimized using the DRL method. Numerical results validate that our JSCE scheme achieves superior semantic throughput compared to the conventional schemes and efficiently reduces the semantic encoder's mode size in multi-user scenarios.
\end{abstract}

% Note that keywords are not normally used for peerreview papers.
\IEEEpeerreviewmaketitle
\let\thefootnote\relax\footnotetext{ The work of Shimin Gong was supported in part by National Natural Science Foundation of China under Grant 62372488 and the Shenzhen Fundamental Research Program under Grant JCYJ20220818103201004. 
The work of Lanhua Li was supported in part by the National Natural Science Foundation of China under Grant 62202506 and Grant 62371478.
%The work of Lanhua Li was supported in part by the National Natural Science Foundation of China under Grant 62202506 and the Guangdong University Featured Innovation Program Project (No. 2023KTSCX004). 
The work of Songhan Zhao was supported in part by the China Scholarship Council. (Corresponding author: Shimin Gong)

%Haidong Wang, Songhan Zhao, Shimin Gong, Bo Gu,and Lanhua Li are with the School of Intelligent Systems Engineering, Sun Yat-sen University, Shenzhen Campus, China, and also with Guangdong Provincial Key Laboratory of Fire Science and Intelligent Emergency Technology, Guangzhou 510006, China (e-mail: wanghd7@mail2.sysu.edu.cn; zhaosh55@mail2.sysu.edu.cn; gongshm5@mail.sysu.edu.cn; gubo@mail.sysu.edu.cn; lilh65@mail.sysu.edu.cn).
}
\vspace{-0.2cm}
\section{Introduction}
\vspace{-0.1cm}
% Shannon's information theory laid the foundation of modern communications. However, with the rapid growth of data in the information age, channel capacities are approaching Shannon's limits. This underscores an urgent need to explore alternative paradigms beyond the Shannon framework to meet soaring communication demands. Semantic communications, which focuses on conveying the meaning rather than the exact form of information, has emerged as one promising approach to overcoming Shannon's bounds. 

% Recently, many deep learning (DL) based joint source-channel coding (JSCC) paradigms have been proposed, which is refered as DeepJSCC and it formulates semantic communication as an end-to-end (E2E) propagation process\cite{Tung2022DeepJSCC, Kurka2020DeepJSCC}. Through emerging DL models and E2E training, we are able to deploy the prior knowledge required for semantic transmission at both the transmitter and receiver ends. And prior knowledge allows us to further compress the data that needs to be transmitted.

%The growing demands for communication highlight the need to explore alternatives to the Shannon framework \cite{Letaief2019_6G}. The 
Semantic communication, which focuses on conveying the meaning of information rather than raw data, has emerged as a promising approach to address Shannon limits such as increasing traffic demand and lower latency requirements \cite{Gündüz2023semcom}. Semantic communication \textcolor{black}{extracts semantic features from raw data, relying on the shared prior knowledge between the transmitters and receivers.}
Typically, this prior knowledge can be obtained by conventional methods like knowledge graphs \cite{Zhao2023graph} or represented in a well-trained encoder-decoder pair using deep joint source-channel coding (DeepJSCC) architecture. The DeepJSCC treats semantic communication as an end-to-end (E2E) system, leveraging both source signal and channel characteristics to achieve higher transmission efficiency, lower complexity, and greater robustness \cite{Bourtsoulatze2019DeepJSCC}. 
% \textcolor{black}{The DeepJSCC also demonstrates greater robustness and advantages such as avoiding the cliff effect.} %\cite{Yilmaz2023distributed}.}

Semantic communication provides a new dimension for multi-user orthogonal channel access, as the semantic feature vectors of different users can be exploited and extracted to be mutually orthogonal \cite{Zhang2023DeepMA}.
Existing work has combined semantic communication with conventional \textcolor{black}{multiple access methods} like non-orthogonal multiple access (NOMA) \cite{Li2023NOMA-Semantic} and rate splitting multiple access (RSMA) \cite{Cheng2023RSMA-semantic-interest_}\textcolor{black}{, demonstrating improved performance in wireless networks.}
However, existing multiple access schemes inadequately exploit both the semantic source and physical channel characteristics. In densely deployed wireless networks with increasing user numbers, multiple access methods based solely on channel characteristics or coding multiplexing become insufficient to effectively serve all users.
To tackle this, some research has investigated the integration of DeepJSCC into multiple access schemes \cite{Zhang2023DeepMA}. The deep learning-based multiple access (DeepMA) \cite{Zhang2023DeepMA} has been shown to outperform conventional communication methods in high signal-to-interference plus noise ratio (SINR) ratio environments, maintaining stable performance even as the number of users increases.  
In densely deployed networks, in addition to channel configuration in the physical layer, the signal's semantic features can be exploited to create orthogonality for simultaneous transmissions. 
This insight motivates us to design a joint \textcolor{black}{source-channel encoding scheme by leveraging both the semantic and spatial features} to improve the spectrum efficiency in multi-user wireless networks.

The intelligent reflecting surface (IRS) has emerged as a promising technique to enlarge our capability for channel configuration in favor of multi-user access. The IRS is composed of a large number of passive reflecting elements. Each element in the IRS is capable of inducing phase shifts on incident signals \cite{2020WuIRS-magazine, Wang2022IRS-DRL}. The author in \cite{Hou2020IRS-NOMA} studied an IRS-assisted NOMA system, showing that the IRS can enhance or reduce channel diversity to improve multi-user services. The author in \cite{Song2021IRS-PLS} explored the physical layer security of a multi-user NOMA network, finding that optimizing IRS beamforming improves secrecy performance.
Inspired by physical layer key generation techniques \cite{2019WuIRS-joint-beamforming}, \textcolor{black}{the IRS can not only enhance the channel conditions for wireless transmissions but also modify the channel state information (CSI), serving as spatial features to facilitate multi-user semantic decoding}. \textcolor{black}{Motivated by this, we leverage both the semantic features of the users' information and the spatial features provided by the IRS-controlled CSI to design a unified encoder-decoder for multiple users.}% We employ it for joint source and channel encoding to improve the performance of high-density multi-user semantic communication systems.
% To utilize IRS as codecs in a bit-oriented scheme, complex coding algorithms are needed due to the wireless environment's complexity.

% we consider that
In this paper, we propose an IRS-assisted joint source-channel encoding (JSCE) scheme that takes advantage of both semantic communication and IRSs.
We extract semantic features from raw images at the source and use the IRS to modify the CSI, providing additional spatial features for multi-user semantic decoding. In particular, we adopt an attention mechanism to merge the IRS-controlled CSI into semantic features, amplifying certain dimensions while suppressing others to achieve higher orthogonality among users' information.
We formulate an optimization problem to maximize the multi-user semantic throughput by jointly optimizing the users' scheduling, IRS's passive beamforming, and semantic encoding strategies. 
\textcolor{black}{To tackle the non-convex problem, we propose an explainable deep neural network-driven deep reinforcement learning (XD-DRL) framework. This framework incorporates a DNN-based semantic encoder for semantic feature extraction from raw images, with IRS phase shifts integrated into the DNN's neurons.  After training, certain DNN weight coefficients become meaningful, representing the optimized IRS passive beamforming.} 
%Due to the non-convexity of the problem, we propose a hierarchical deep reinforcement learning (DRL) framework that decomposes the original problem into two steps. First, the users' scheduling strategy is adapted by the outer-loop deep deterministic policy gradient (DDPG) method. 
Then, given the optimized \textcolor{black}{IRS passive beamforming and semantic encoding strategies, we employ the deep deterministic policy gradient (DDPG) method to adapt the users' scheduling.} 
Numerical results validate that our proposed JSCE scheme achieves higher throughput performance compared to benchmark methods and significantly reduces the model size in multi-user scenarios.

%\vspace{-0.1cm}
\section{System model}
%\vspace{-0.1cm}

\begin{figure}
    \centering
    \includegraphics[width=0.42\textwidth]{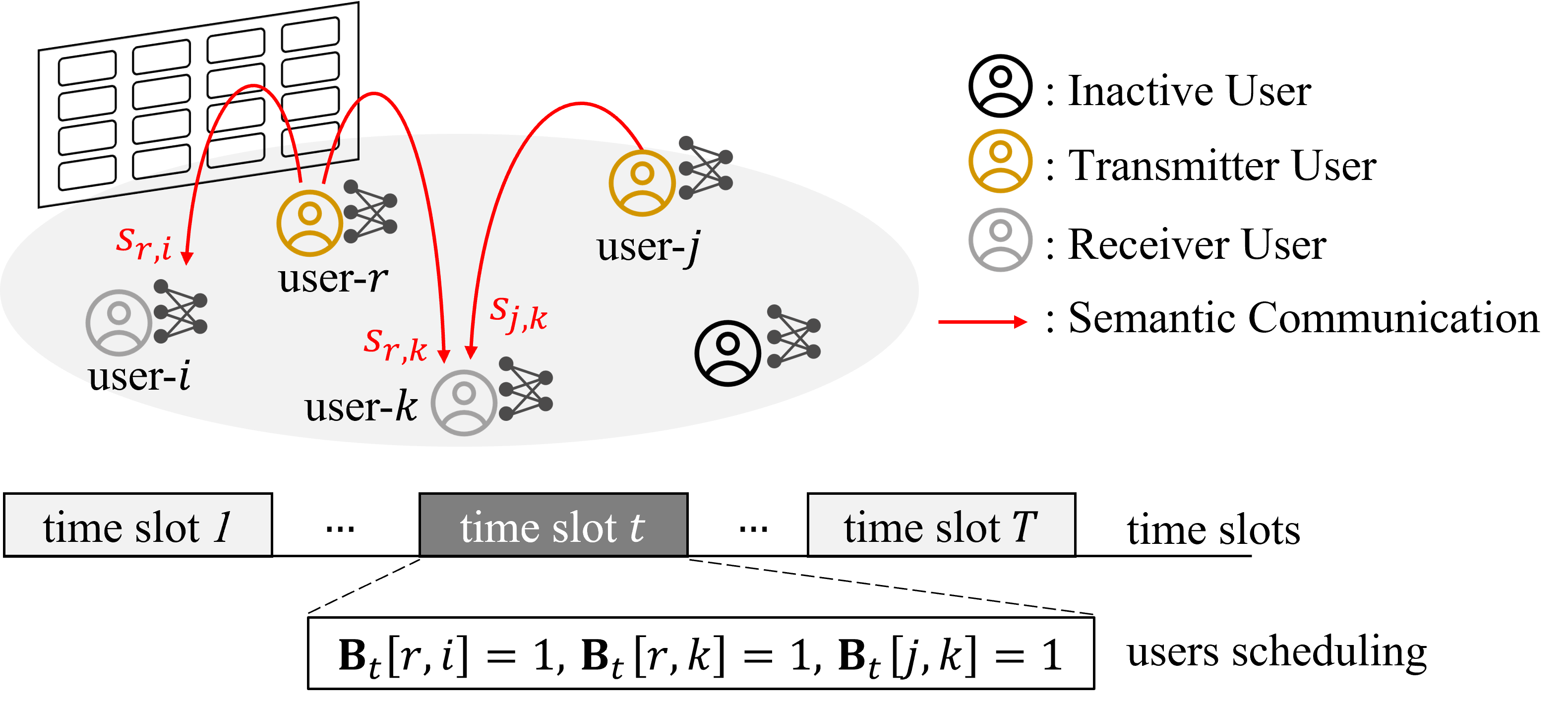}
    \caption{The IRS-assisted multi-user semantic communication. }
    \label{fig:system_model}
    \vspace{-0.5cm}
\end{figure}

As illustrated in Fig.~\ref{fig:system_model}, we consider a semantic-aware and IRS–assisted multi-user wireless network.
%\textcolor{black}{which supports concurrent point-to-point, point-to-multiple, and multiple-to-point transmissions simultaneously}. 
% We assume that there are $K$ single-antenna half-duplex users and the $k$-th user is denoted as user-$k$.
The set of users is denoted as $\mathcal{K}=\{1,\ldots, K\}$. Each user is equipped with a semantic encoding unit for extracting semantic information from raw data. We denote the $k$-th user as user-$k$ and the direct channel from user-$k$ to user-$r$ as $h_{k,r}$. We consider that all users' channels are reciprocal, i.e., $h_{k,r}=h_{r,k}$.
We assume a time-slotted transmission protocol, as shown in Fig.~\ref{fig:system_model}. The users' scheduling strategy in the time slot-$t$ is represented as an adjacent matrix $\mathbf{B}_{t}\in\{0,1\}^{K\times K}$. Let $\mathbf{B}_{t}[r,k]=1$ represent that the user-$r$ communicates with the user-$k$ in the $t$-th time slot. The semantic information from user-$r$ to user-$k$ is denoted as $s_{r,k}$. The IRS with $N$ reflecting elements can improve all users' channel conditions by inducing passive beamforming in the wireless communication system.
%\vspace{-0.1cm}
\subsection{IRS-assisted Channel Model}
We consider that the channel $\mathrm{g}_k$ from the user-$k$ to the IRS follows the Rican channel model as follows:
\begin{equation}
    \mathrm{g}_{k} =\sqrt{\frac{K}{K+1}} \mathrm{g}_{k, \text{LoS}} + \sqrt{\frac{1}{K+1}} \mathrm{g}_{k,\text{NLoS}},
    \label{equ:rician channel}
\end{equation}
\vspace{0.2cm}
where $\mathrm{g}_{k,\text{LoS}}$ is the line-of-sight (LoS) component of $\mathrm{g}_k$ and can be represented as follows:
\begin{equation}
    \mathrm{g}_{k, \text{LoS}} = e^{-k_0d_k}\cdot \mathbf{a}(\varphi_{k} ,\theta_{k} ),
    \label{equ:channel_k}
\end{equation}
where $d_k$ is the distance between the user-$k$ and the IRS. The exponent $-k_0d_k$ denotes the phase shift over propagation, where $k_0=\frac{2\pi}{\lambda}$ is determined by the wavelength of the signal $\lambda$. The fast-fading non-LoS component $\mathrm{g}_{k,\text{NLoS}}$ is a complex Gaussian random variable.
The IRS is modeled as a uniform planar array (UPA) with the array response $\mathbf{a}(\varphi, \theta)$ derived using the Saleh-Valenzula (SV) channel model, which is a function of elevation angle $\theta$ and azimuth angles $\varphi$. To avoid grating lobes, the interval of elements is set to half of the wavelength, i.e., $\lambda/2$. Thus, the channel response is given by:
\begin{equation}
    \begin{aligned}
        \mathbf{a}(\varphi ,\theta ) = & [1,...,{e^{j\pi(m\sin \varphi \cos \theta  + n\sin \theta )}}, \\ & ...,{e^{j\pi((N - 1)\sin \varphi \cos \theta  + (N - 1)\sin \theta )}}], \notag
    \end{aligned}
%    \vspace{-0.1cm}
\end{equation}
where $m$ and $n$ are the indices of the IRS reflecting elements.
\textcolor{black}{We define $\mathbf{\Phi}\in \mathbb{C}^{N\times N}$ as the reflection matrix}, and the channel $\mathbf{h}_{r,k}$ from user-$r$ to user-$k$ is represented as follows:
\begin{equation}
\mathbf{h}_{r,k}=\mathrm{g}_{k}\mathbf{\Phi}\mathrm{g}^{H}_{r} + h_{r,k}.
    \label{equ:composite channel}
%    \vspace{-0.1cm}
\end{equation}

\vspace{-0.4cm}
\subsection{\textcolor{black}{Joint Source-Channel Encoding}}
Let $w_{r,k}$ represent the raw data transmitted from user-$r$ to user-$k$. Defining the adjustable parameters $\theta_s, \theta_c$, the semantic and channel encoders are represented by $\mathcal{SE}_{\theta_s}$ and $\mathcal{CE}_{\theta_c}$, respectively. Thus, the semantic features are written as follows:
\begin{equation}
    s_{r,k}=\mathcal{SE}_{\theta_s}(w_{r,k}).\label{equ:channel_encoding}
\end{equation}
To ensure successful decoding of each user's signal, we incorporate each user's CSI into the codebook. 
The semantic feature incorporating CSI can be expressed as follows:
\begin{equation}
s^a_{r,k} =  a_{r,k}\odot s_{r,k},
\end{equation}
where $a_{r,k} = \mathcal{CE}_{\theta_c}(\mathbf{h}_{r,k},w_{r,k})$ and operator $\odot$ is the Hadamard product.
\textcolor{black}{For brevity}, we merge the source coding and channel coding as a JSCE encoder $\mathcal{E}_{\theta}(\mathbf{h}_{r,k}, w_{r,k})$, where $\theta=\{\theta_s, \theta_c\}$ contains parameters for both the source and the channel encoder.
\textcolor{black}{Note that $\mathcal{E}_{\theta}$ is user-independent, which is consistent for 
%\vspace{0.2cm}
all users.} If user-$r$ tends to communicate with multiple users, the semantic feature sent can be denoted as:
\begin{equation}
    s_r=\sum_{j=1}^{K}{\mathbf{B}_t[r,j] s^a_{r,j}}.
    \label{equ:overlapping_with_csi_embedding}
\end{equation}
Note that the semantic feature transmitted by the user-$r$ is normalized to ensure that its signal satisfies the power constraint.

\subsection{Semantic Decoding}
We assume that the transmission of an entire semantic feature can be accomplished within a channel coherence time. Thus, the signal received at user-$k$ can be written as follows:
\begin{equation}
    y^{(i)}_{k}=\sum^{K}_{r=1}\mathbf{B}_{t}[r,k]\mathbf{h}_{r,k}{\sum^{K}_{j=1}\mathbf{B}_t[r,j]s^{a(i)}_{r,j}}+n^{(i)}_{r,k}, i\in \{1,...,L\},
    \label{equ:propagation process}
\end{equation}
where $n^{(i)}_{r,k}$ is the Gaussian noise, $i$ is the index of the $i$-th semantic symbol, and $L$ denotes the length of the semantic information vector. In summary, the received signal undergoes two superpositions: the first appears during semantic encoding at the source and the second appears during the propagation in the wireless channel controlled by the IRS.

 We represent the decoder as an inverse function with parameters $\tilde{\theta}$. \textcolor{black}{The decoded data can be written as follows:}
\begin{align}
\hat{w}_{r,k}=\mathcal{E}_{\tilde{\theta}}^{-1}\left(\hat{s}_k, \mathbf{h}_{r,k}\right), 
\label{equ:decode}
\end{align}
where $\hat{s}_k=[y^{(1)}_{k},...,y^{(L)}_{k}]\in  \mathbb{C}^{L}$ denotes the received semantic vector. \textcolor{black}{The SINR from user-$r$ to user-$k$ is denoted as follows:}
\begin{equation}
    \gamma_{r,k} = \frac{P_t|\mathbf{h}_{r,k}|^2}{\sigma^2+I^e_{r,k}+I^t_{r,k}},
    \label{equ:SNR}
\end{equation}
where $\sigma^2$ denotes the noise power and $P_t$ is the normalized transmit power for all users. 
The interference $I^e_{r,k}$ and $I^t_{r,k}$ received at the user-$k$ arises from both the encoding and transmission processes, represented as follows:
\begin{align}
    & I^e_{r,k} =  \mathbb{E}[|\mathbf{h}_{r,k}\sum^K_{\substack{i=1,i\neq k}}\mathbf{B}_t[r,i]s^a_{r,i}|^2], \notag \\ 
    & I^t_{r,k} =  \mathbb{E}[\sum^K_{\substack{j=1,j\neq {r}}}\mathbf{B}_t[j,k]|\mathbf{h}_{j,k} s_j|^2].
\end{align}
Note that the joint source-channel encoding can reduce the users' interference from both $I^e_{r,k}$ and $I^t_{r,k}$.   
%The interference $I^c_{r,k}$ means user-$r$ overlaps semantic signals for users from $1$ to $K$ but user-$k$ demand $s_{r,k}$ only. $I^t_{r,k}$ means that not only is user-$r$ transmitting but also other users. $x_j$ denotes the overlapped signal from user-$j$ according to (\ref{equ:propagation process}). The interference can be represented as follows:
%\vspace{-0.3cm}

\subsection{Semantic Throughput}
To evaluate the transmission performance of the semantic communications, we define the semantic throughput (measured in semantic units (suts), similar  to that in~\cite{Li2023NOMA-Semantic, Yan2022SemRate4text}) as follows:
% We define the image size as $I_I=H\times W\times 3$, and the size of the semantic features is $L$. The semantic transmission metric named semantic unit per second is defined as:
\begin{equation}
    \Gamma_{r,k} = \frac{BS}{C_rI}\xi(\gamma_{r,k},\theta,\tilde{\theta})~,
\end{equation}
where $S$ is the average semantic information carried in the image, coefficient $C_r$ is the compression ratio, $B$ is the channel bandwidth, \textcolor{black}{and $I$ represents the number of bits of raw data}. 
In this paper, we assume $S=\mathcal{M}_{\text{tr}}(I)$, where $\mathcal{M}_{\text{tr}}$ denotes a traditional modulation scheme that maps data from bits to symbols. Meanwhile, we have $C_rI=L$ for the proposed JSCE scheme. 
The semantic similarity $\xi(w_{r,k}, \hat{w}_{r,k})$ is derived based on the difference between $w_{r,k}$ and $\hat{w}_{r,k}$. The received data $\hat{w}_{r,k}$ is an implicit function of SINR $\gamma_{r,k}$ and the semantic encoding parameters $\{\theta, \tilde{\theta}\}$.
As such, the semantic similarity can be formulated using structure similarity (SSIM) as follows:
\begin{equation}
    \xi(w_{r,k}, \hat{w}_{r,k}) = \frac{(2\mu_{w_{r,k}}\mu_{\hat{w}_{r,k}}+c_1)(2\sigma_{w_{r,k}\hat{w}_{r,k}}+c_2)}{(\mu^2_{w_{r,k}}+\mu^2_{\hat{w}_{r,k}}+c_1)(\sigma^2_{w_{r,k}}+\sigma^2_{\hat{w}_{r,k}}+c_2)},
    \label{equ:ssim}
\end{equation}
where $\mu$ denotes the pixel sample mean value and $\sigma^2$ is the variance. The coefficient $\sigma_{w\hat{w}}$ is the covariance of $w_{r,k}$ and $\hat{w}_{r,k}$, and $c_1, c_2$ are the constants that stabilize the division with a weak denominator. 
% Note that the SINR $\gamma_{r,k}$ affects $\hat{w}_{r,k}$ indirectly as shown in (\ref{equ:propagation process}). Thus, we denote $\hat{w}_{r,k}$ as a function of $\gamma_{r,k}$ and semantic encoding parameters $\{\theta,\tilde{\theta}\}$. 
The decoded data $\hat{w}_{r,k}$ is determined by the semantic encoding $\{\theta, \tilde{\theta}\}$ and the SINR $\gamma_{r,k}$.
Finally, the semantic throughput $\Gamma_{r,k}$ from user-$r$ to user-$k$ can be reformulated as follows:
\begin{equation}
    \Gamma_{r,k} = \frac{B\mathcal{M}_{\text{tr}}(I_I)}{L}\xi\left(w_{r,k}, \hat{w}_{r,k}|(\gamma_{r,k}, \theta, \tilde{\theta})\right).
    \label{equ:gamma}
\end{equation}
The SINR $\gamma_{r,k}$ depends on the number of access users and current channel conditions. To improve semantic throughput, the semantic encoding parameters $\{\theta, \tilde{\theta}\}$ must be fine-tuned based on the current interference level.

\section{\textcolor{black}{Explainable DNN-driven Learning} for Semantic Throughput Maximization}
% 这段需要重新，你需要在提出你的优化问题前，先把各个变量之间的关系写清楚，为什么要联合优化这几个变量，比如：语义提取可以增加用户信息的语义正交性，反射面可以增加用户的信道正交性，分别是怎么促进用户的信息传输和恢复的。巴拉巴拉介绍很多。然后说因此我们可以通过联合优化语义和irs来研究这个优化问题。
% Given a period of time, 
Considering the fairness among users, we maximize the minimum semantic throughput of the multiple users by jointly optimizing the users' scheduling $\mathbf{B}_t$, the IRS's passive beamforming $\mathbf{\Phi}$, and the semantic encoding $\{\theta, \tilde{\theta}\}$. 
% \textcolor{red}{Note that the semantic encoding improves the information orthogonality of all users' signals, while the IRS's passive beamforming can improve the channel orthogonality.}
We formulate the max-min optimization problem as follows:
\begin{subequations}\label{equ:main_problem}
\begin{align}
\max_{\mathbf{\Phi},\mathbf{B},\theta,\tilde{\theta}}&~\min_{r,k\in\mathcal{K}}\frac{1}{T}\sum_{t=1}^{T}\mathbf{B}_t[r,k]\xi\left(w^{(t)}_{r,k}, \hat{w}^{(t)}_{r,k}|(\gamma^{(t)}_{r,k}, \theta, \tilde{\theta})\right) \tag{\ref{equ:main_problem}} \\
\mathrm{s.t.}~&~(\ref{equ:channel_encoding})-(\ref{equ:gamma}), \\
    ~&~|\varphi_{n}|\leq 1, \deg\left(\varphi_n\right)\in\{0,\pi\}, \forall{n}\in [1, N], \label{constraint:IRS}\\
    ~&~\sum^{K}_{k=1}\mathbf{B}_t[k,r]=0, \forall{r}\in\mathcal{K}_{T}(t), \label{constraint:half_duplex}
\end{align} 
\end{subequations}
where $\mathcal{K}_{T}(t)\subseteq \mathcal{K}$ denotes the \textcolor{black}{transmitter set} at $t$-th time slot. \textcolor{black}{The IRS's passive beamforming $\mathbf{\Phi}$ is represented by $\mathbf{\Phi}=\text{diag}([\varphi_1,...,\varphi_N])$, where $\varphi_n$ is the phase shift induced by the $n$-th IRS reflecting element.} Constraints~\eqref{constraint:IRS} defines the 1-bit IRS's reflection capacity, while constraint~\eqref{constraint:half_duplex} ensures that each user operates in half-duplex. Problem (\ref{equ:main_problem}) is difficult to solve directly due to the lack of an explicit expression between $\{\theta, \tilde{\theta}\}$ and $\xi$. \textcolor{black}{Note that the successful decoding of multi-user's semantic information requires ensuring both semantic and spatial orthogonality among the users.}  

\begin{figure}
    \centering
    \includegraphics[width=0.46\textwidth]{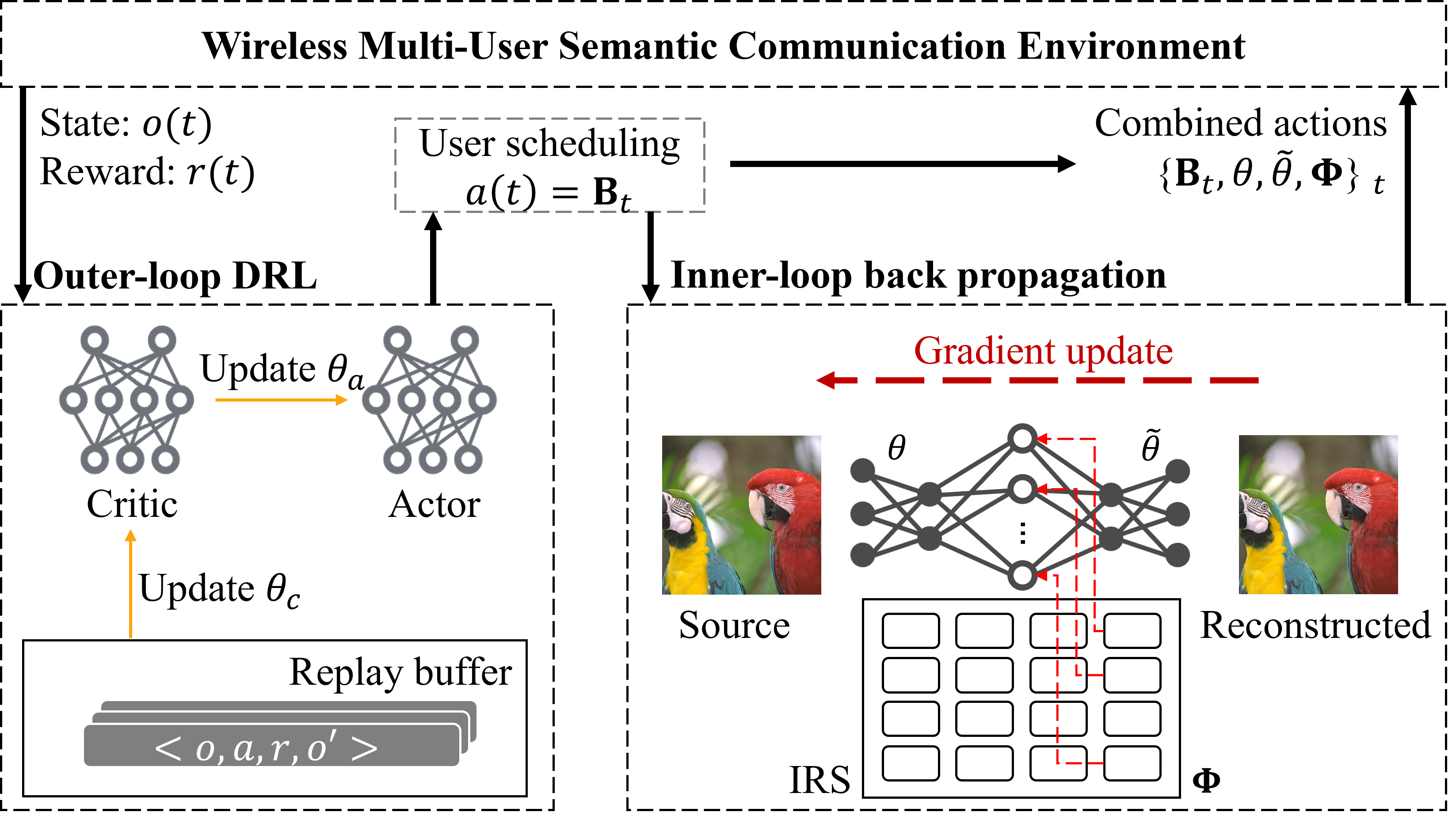}
    \caption{The proposed \textcolor{black}{XD-DRL} framework.}
    \label{fig:algorithm_framework}
    \vspace{-0.45cm}
\end{figure}

%\textcolor{black}{The semantic similarity $\xi$ is monotonically increasing with respect to SINR \cite{Yan2022SemRate4text}. Thus,}
To solve this complex problem, we decompose problem (\ref{equ:main_problem}) into two stages, i.e., maximizing users' minimal SINR and maximizing semantic similarity. We then solve these stages alternatively. 
%However, $\mathbf{\Phi}, \theta, \tilde{\theta}$ and $\mathbf{B}$ are not decoupled completely due to the limited control capability of the IRS. Thus, we optimize problem (\ref{equ:schedule_problem}) and (\ref{equ:semantic_problem}) alternatively. 
We propose a two-step \textcolor{black}{XD-DRL} algorithm, as shown in Fig.~\ref{fig:algorithm_framework}. The algorithm includes the outer-loop DRL for the users' scheduling strategy and the inner-loop back-propagation training for the semantic encoding and the IRS's passive beamforming strategies. In each time slot, the DRL first outputs the users' scheduling strategy $\mathbf{B}_t$. Given $\mathbf{B}_t$, the IRS's passive beamforming and semantic encoding are jointly optimized with backpropagation.

\subsection{Explainable Learning for IRS's Passive Beamforming and Semantic Encoding Strategies}
Given the users' scheduling strategy $\mathbf{B}_t$, we focus on maximizing the semantic similarity by jointly optimizing the IRS's passive beamforming $\mathbf{\Phi}$ and the semantic encoding parameters $\{\theta,\tilde{\theta}\}$, and the subproblem can be written as follows:
\begin{subequations}\label{equ:semantic_problem}
\begin{align}
\max_{\mathbf{\Phi}, \theta, \tilde{\theta}}&~{\xi\left(w_{r,k}, \hat{w}_{r,k}|(\gamma_{r,k}, \theta, \tilde{\theta})\right)} \tag{\ref{equ:semantic_problem}} \\
\mathrm{s.t.} ~&~(\ref{equ:channel_encoding})-(\ref{equ:decode}), \\
    ~&~|\varphi_{n}|\leq 1, \deg\left(\varphi_n\right)\in\{0,\pi\}, \forall{n}\in [1, N].
\end{align} 
\end{subequations}

To solve problem (\ref{equ:semantic_problem}), a DNN is developed to map the encoding and decoding processes. The DNN backbone is referenced from \cite{Xu2022image_attention, Zhang2023DeepMA}, consisting of basic residual blocks (BRB), inverted basic residual blocks (IBRB), and channel attention blocks (CAB).
%The detailed encoder structure is described in Fig.~\ref{fig:JSCE_Original}. 
% Each basic residual block (BRB) module is equivalent to a 2D convolution with a padding of 1 and a kernel size of 3x3 as shown in Fig.~\ref{fig:JSCE_BRB}. 
The IBRB module is achieved by replacing the convolution module with a transposed convolution. The CSI is merged into the semantic feature with the CAB. %in \cite{Xu2022image_attention}. 

During encoding, we first embed the scalar or low-dimensional CSI into a high-dimensional vector space using a CSI-to-vector (C2V) mapping function. The C2V function maps similar CSIs to proximate positions in the embedding space. For simplicity, we consider a $2$-D environment with all users on the same plane. Following the method in \cite{vaswani2023attention}, we construct a $2$-D position embedding denoted as $e_k=\mathrm{C2V}(\mathrm{g}_{k})$. We then compute the channel-wise mean vector of the input feature map and add it to the CSI embedding $e_k$. Using a multi-layer perceptron (MLP) and a softmax output layer, we calculate the channel attention $a_{r,k}\in \mathbb{C}^L$ as follows:
\begin{equation}
    a_{r,k} = \text{softmax}(\text{MLP}(s_{r,k}+e_k)).
    \label{equ:attention}
\end{equation}
\textcolor{black}{The IRS improves all users' spatial orthogonality by the passive beamforming, as shown in (\ref{equ:composite channel}). To optimize the IRS's passive beamforming, we integrate the IRS's phase shift $\mathbf{\Phi}$ into the DNN architecture. We assign certain weight coefficients of the DNN to represent the IRS's phase shifts and optimize them using the back-propagation method. After training, we apply the quantization method to map the IRS's phase shift to the range $\{0,\pi\}$, ensuring compliance with constraint (\ref{constraint:IRS}). Thus, the IRS's beamforming and semantic encoding strategies can be jointly optimized by training a DNN-based encoder.}
%We define a trainable parameter $\mathbf{W}_{r}\in \mathbb{R}^N$, corresponding to the phase angle of the IRS. Gradient updates will be applied to $\mathbf{W}_{r}$ rather than $\mathbf{\Psi}$, constraining the magnitude of the reflection coefficients and the magnitude of phase variation. After the training is complete, we quantize the values in $\mathbf{W}_{r}$ to $\{0,\pi\}$.}

We maximize the semantic similarity by minimizing the mean square error (MSE). The training process is similar to the autoencoder. The input image serves as the ground truth label, making it a self-supervised E2E training. Given the scheduling policy, we calculate the MSE loss between the recovered image at the receiver and the original image at the transmitter.

\subsection{DDPG for Users' Scheduling Strategy}
Given the IRS's passive beamforming and semantic encoding, we can rewrite the users' scheduling optimization subproblem by replacing the implicit function $\xi$ with the SINR in the $t$-th time slot $\gamma^{(t)}_{r,k}$ as follows: 
\begin{subequations}\label{equ:schedule_problem}
\begin{align}
\max_{\mathbf{B}}&~\min_{r,k\in\mathcal{K}}\frac{1}{T}\sum_{t=1}^{T}\mathbf{B}_t[r,k]\gamma^{(t)}_{r,k} \tag{\ref{equ:schedule_problem}}\\
\mathrm{s.t.} ~&~(\ref{equ:SNR})-(\ref{equ:gamma}), \\
~&~\sum^{K}_{k=1}\mathbf{B}_t[k,r]=0, \forall{r}\in\mathcal{K}_{T}(t).
\end{align} 
\end{subequations}
\textcolor{black}{The optimization of the scheduling strategy defines a discrete feasible set according to constraint (\ref{constraint:half_duplex}), which is challenging to solve.} Therefore, we employ the DDPG method to solve problem (\ref{equ:schedule_problem}). We first reformulate the users' scheduling into a Markov decision process (MDP) as follows:
\begin{enumerate}
    \item \textbf{State:} 
    \textcolor{black}{The state $o(t)\in  \mathbb{N}^{K}$ is defined as the schedule history of each user up to time $t$, i.e., $o(t)[j] = o(t-1)[j] + \sum^{K}_{i\neq j}\mathbf{B}_{t-1}[j,i] + \sum^{K}_{i\neq j}\mathbf{B}_{t-1}[i,j]$  
    }
    %The state observation is defined as $o(t)=\eta_t$, where $\eta_t\in \mathbb{N}^{K}$ represents the schedule history of each user up to time $t$. For example, $\eta_{t+1}[i] = \eta_t+1$ if user-$i$ is scheduled in time slot-$t$ regardless of whether user-$i$'s function is transmitting or receiving information.
    \item \textbf{Action:} The action at the $t$-th time slot is defined as the users' scheduling strategy $\mathbf{B}_t$. 
    \item \textbf{Reward:} The instantaneous reward at each time instant is defined as the accumulated minimum SINR from the start of scheduling to the current time $t$, as follows:
    \begin{equation}
r(t)\!=\!\min_{r,k\in\mathcal{K}}\sum_{i=1}^{t}\mathbf{B}_i[r,k]\xi\left(w^{(i)}_{r,k}, \hat{w}^{(i)}_{r,k}|(\gamma^{(i)}_{r,k}, \theta, \tilde{\theta})\right)\!.
        \label{equ:reward}
         \vspace{-0.1cm}
    \end{equation}
\end{enumerate}

The DDPG method uses a set of online networks with parameters of $\theta_a, \theta_c$, and a set of target networks with parameters of $\theta'_a, \theta'_c$ to stabilize the training. The subscripts $a$ and $c$ denote the actor network and the critic network, respectively. Given the current state $o(t)$, the objective of DRL is to generate an action $a(t)=\pi(o(t)|\theta_a)$ to maximize the value function $J(\theta_a)$. For the scheduling problem, we define the value function as the expected value of the action-value function, evaluated by the critic network with parameter $\theta_c$, which can be expressed as:
\begin{equation}
    J(\theta_a)\approx \mathbb{E}_{B}[Q(o,a|\theta_c)],
\end{equation}
where the Q-value $Q(o,a|\theta_c)$ is the output of the critic and the subscript $B$ denotes the mini batch sampled from reply buffer. Then, the actor network is updated by using the deterministic policy gradient as follows:
\begin{equation}
    \nabla_{\theta_a} J =  \mathbb{E}_{B}[\nabla_{\theta_a} \pi \nabla_{a} Q(o,a|\theta_c)]. \label{equ:policy gradient}
     \vspace{-0.1cm}
\end{equation}
We define $y_t$ as the Q-value, which is calculated as follows:
\begin{equation}
    y_i = r_i+\lambda Q(o_i,\pi(o_i|\theta'_a)|\theta'_c).
\end{equation}
To minimize the TD error, the online critic $\theta_c$ is updated with:
\begin{equation}
    \mathcal{L}_c =  \mathbb{E}_{B}[(y_i-Q(o,a|\theta_c))^2].\label{equ:critic loss}
    \vspace{-0.1cm}
\end{equation}
% After certain steps of BP, the target network will be updated with online network.

The details of the proposed \textcolor{black}{XD-DRL} algorithm are summarized in Algorithm~\ref{alg:DRL train}. The maximum number of episodes and the steps per episode are denoted as $E$ and $S$, respectively. The parameter $R_f$ represents the replay frequency, and $U_f$ denotes the update frequency. It is worth noting that the backpropagation training in line~\ref{line:semantic training} is time consuming, introducing significant training costs for the overall DRL. To address this, we pre-train a set of parameters $\{\theta^g, \tilde{\theta}^g\}$ by considering that all users simultaneously communicate to each other, i.e., $\mathbf{B}_t[i,j]=1,\forall{i,j}\in\mathcal{K}$. These serve as initialization to accelerate backpropagation training.

\begin{algorithm}[t] 
\caption{XD-DRL framework for the users’ scheduling, IRS's passive beamforming, and semantic encoding strategies} 
\label{alg:DRL train} 
\begin{algorithmic}[1] %这个1 表示每一行都显示数字
% \REQUIRE ~~\\ %算法的输入参数：Input
\STATE Initialize the number of the user $K$, IRS's size $N$, the users' positions, and the network parameters $\theta, \tilde{\theta}, \theta_a, \theta_c, \mathbf{\Phi}$.
% \ENSURE ~~\\ %算法的输出：Output
% $\theta, \tilde{\theta}, \theta_a, \theta_c$;

\FOR{$n=1:E$}
\FOR{$t=1:S$}
\STATE Update the users' scheduling $\mathbf{B}_t$ by the actor-network\label{line:action}
\STATE Update the IRS's passive beamforming $\mathbf{\Phi}$ and\\
 semantic encoding $\{\theta,\tilde{\theta}\}$ by the backpropagation\label{line:semantic training}
% observe reward
% \STATE Sample $b\times K$ images, denoted as $w=[w_1,...,w_K]$, from validation set;
%\STATE Semantic encoding and transmitting using (\ref{equ:propagation process})
%\STATE Decode semantic feature with (\ref{equ:decode})
\STATE Estimate $r(t)$ by the reward function (\ref{equ:reward})
% \STATE $\eta_{t+1}[i] = \eta_t[i] + 1$ \textbf{if} user-$i$ is scheduled \textbf{else} $\eta_t[i]$; \label{line:update state}
\STATE Update the next state $o(t+1)$
\STATE Store the transition $\{o(t), a(t), r(t), o(t+1)\}$ 

% BP
\IF{$t~\text{mod}~R_f = 0$}
\STATE Sample a mini-batch $B$ from replay buffer
\STATE Update $\theta_c$ and $\theta_a$  by (\ref{equ:critic loss}) and (\ref{equ:policy gradient}), respectively
\ENDIF 

% update model
\IF{$t~\text{mod}~U_f = 0$}
\STATE $\theta'_a = \tau \theta_a + (1-\tau)\theta'_a$, $\theta'_c = \tau \theta_c + (1-\tau)\theta'_c$
\ENDIF 
\ENDFOR
\ENDFOR
\end{algorithmic}
\end{algorithm}
%\vspace{-0.1cm}
\section{Simulation Results}
%\vspace{-0.1cm}
In this section, numerical results are shown to validate \textcolor{black}{the performance of the JSCE scheme in the IRS-assisted multi-user semantic communication systems. We consider $K=5$ users working over $T=5$ time slots.}
% \subsection{Simulation Settings}
% In each transmission experiment, the channel SNR is a hyperparameter that determines the power of the added Gaussian noise, and the noise is generated randomly based on the power. 
The users are uniformly distributed around at (1.13, 0.50), (-0.01, -0.21), (-1.10, -0.28), (0.19, 1.01), (0.20, 0.01), while the IRS is deployed at (0, 0) to provide service to all users. The Rician factor $K$ is set to 10.
\textcolor{black}{We compare the JSCE scheme with four benchmark schemes, i.e., Bit-TDMA, Semantic-TDMA, Semantic-NOMA, and DeepMA.} For Bit-TDMA, we set the length of the LDPC code block $n=1296$ and the rate of the code $R=2/3$. Thus, we can determine the number of bits in the same parity check equation $d_c=(1-R) n=432$ and the number of parity check equations $d_v=144$ using PyLDPC. The virtual channel noise power is fixed as $\sigma^2=0.1$ and 16-QAM is adopted.
All semantic-based multiple access schemes were trained on the CIFAR-10 dataset with 50,000 images and fine-tuned on a subset of ImageNet \cite{Deng2009ImageNet} with 40,000 images. The images for the test are sourced from the Kodak24 dataset, with the peak signal-to-noise ratio (PSNR) used as the evaluation metric. In Semantic-NOMA, scheduling is limited to activating one transmitter at a time, and the successive interference cancellation (SIC) technique is integrated into the DNN-based semantic decoder, as referenced in~\cite{Li2023NOMA-Semantic}. Note that research~\cite{Li2023NOMA-Semantic} focuses on a two-user downlink NOMA transmission, while we allow the number of access users to vary in each time slot.

\begin{figure}
    \centering
    \includegraphics[width=0.38\textwidth]{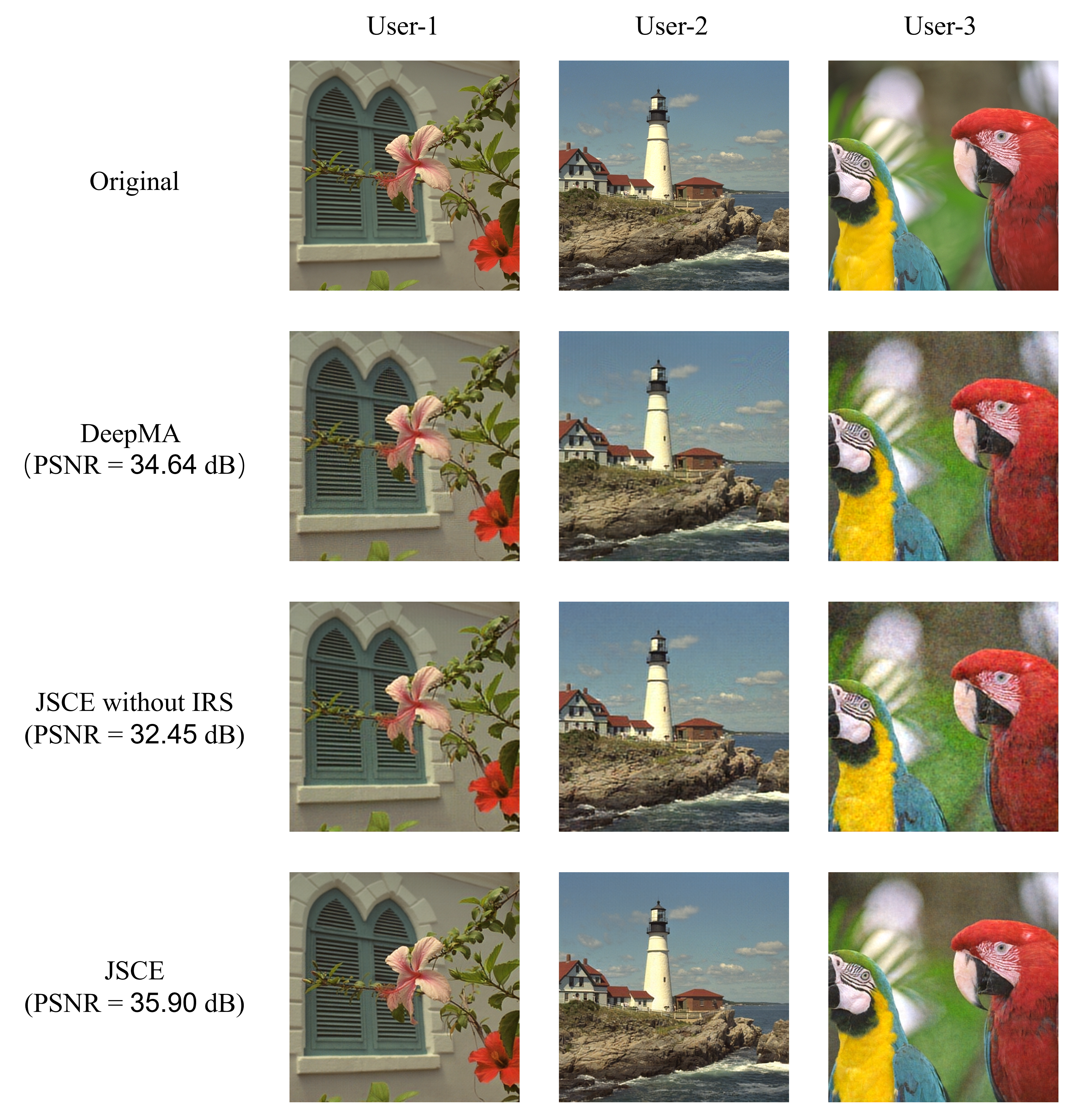}
    \caption{Reconstruction results of the proposed JSCE scheme.}% under slow fading channel. The proposed system transmits 3 pictures corresponding to 3 users simultaneously.}
    \label{fig:JSCE-Original_results}
    \vspace{-0.7cm}
\end{figure}
Figure~\ref{fig:JSCE-Original_results} presents the simulation results where the user at (1.13, 0.50) acts as the transmitter, while the users at (-0.01, -0.21), (-1.10, -0.28), and (0.19, 1.01) are the receivers. The transmitter sends three different $512\times512$ images to these receivers. Fig.~\ref{fig:JSCE-Original_results} demonstrates the feasibility of the proposed JSCE scheme, showing that all users can achieve successful communication using a unified semantic model without additional encoders or decoders. \textcolor{black}{This is because the JSCE scheme employs the IRS to offer distinct spatial features for different users, significantly improving the multi-user's decoding.} The PSNR achieved by the proposed JSCE scheme in this transmission reaches 35.90 dB, comparable to the 36.47 dB attained by DeepMA \cite{Zhang2023DeepMA}. In Fig.~\ref{fig:JSCE-Original_results}, removing the IRS from the JSCE scheme results in a PSNR drop of approximately 3.5 dB, confirming the IRS's significant performance improvement.

\begin{figure}
    \centering
    \includegraphics[width=0.36\textwidth]{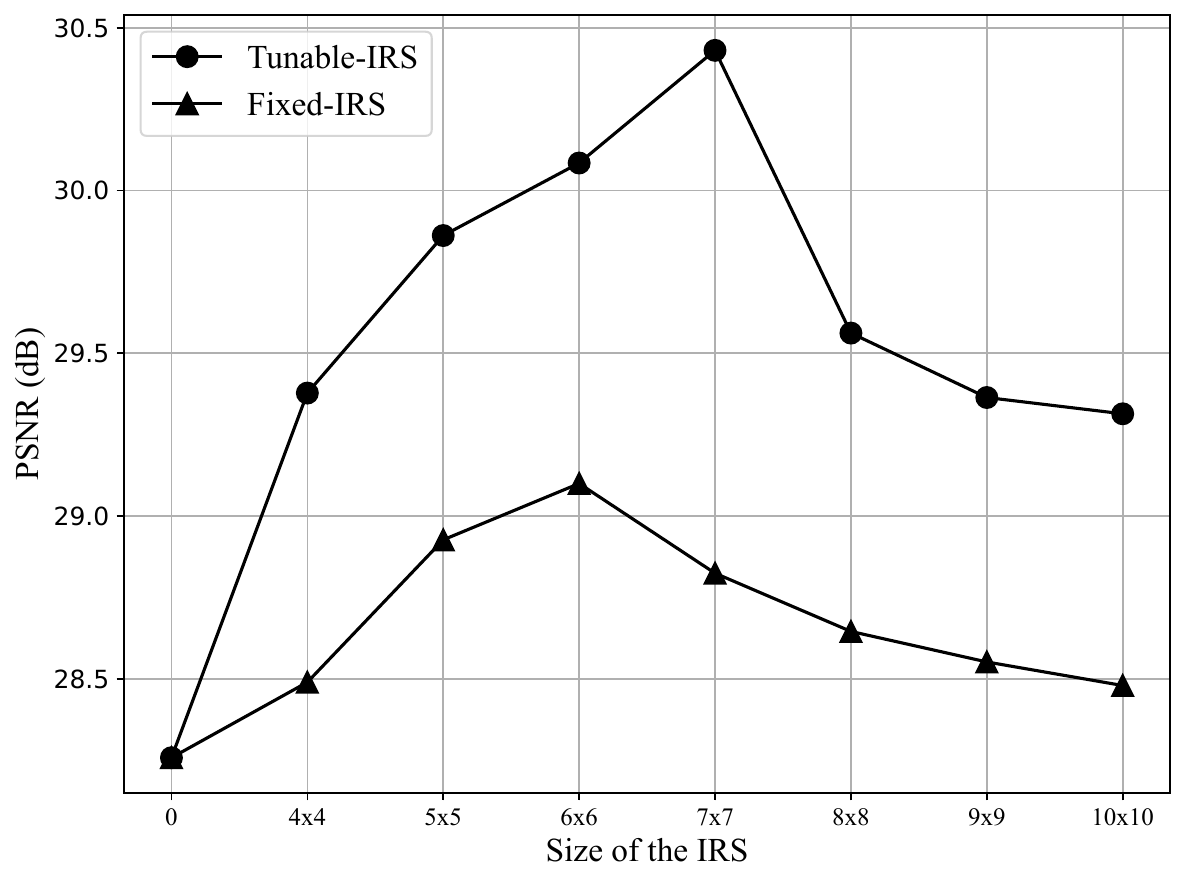}
    \caption{The PSNR performance with varying IRS sizes.}
    \label{fig:IRS-vs-PSNR}
    \vspace{-0.5cm}
\end{figure}

\textcolor{black}{Figure~\ref{fig:IRS-vs-PSNR} reveals the impact of the IRS's size on the JSCE scheme.} \textcolor{black}{We consider a simplified scenario where all users are simultaneously scheduled.} Each user broadcasts a $512\times512$ image and decodes four images from the other users. One group of the IRS's passive beamforming is optimized, while others remain fixed. Initially, the IRSs of all schemes are randomly set to the same state. As the number of the IRS's reflecting elements increases, it is interesting to observe that the PSNR first rises and then decreases. This may be the consequence of the IRS functioning as part of the semantic encoding, similar to a one-layer MLP. Its activation function acts as a periodic function that quantifies the IRS's phase shift to $[-\pi, \pi]$. Optimizing the IRS via backpropagation can lead to gradient issue due to phase discontinuity, causing model training to deviate from the optimal solution. This issue intensifies with larger IRS sizes.
% It is widely believed that larger reflecting surfaces introduce more complex calculations. Therefore, a larger IRS may need longer training to reach its optimal phases. Similar scenarios can be seen in \cite{stylianopoulos2022onlinerisconfigurationlearning}, where a larger IRS resulted in worse performance.

\begin{figure}
    \centering
    \includegraphics[width=0.36\textwidth]{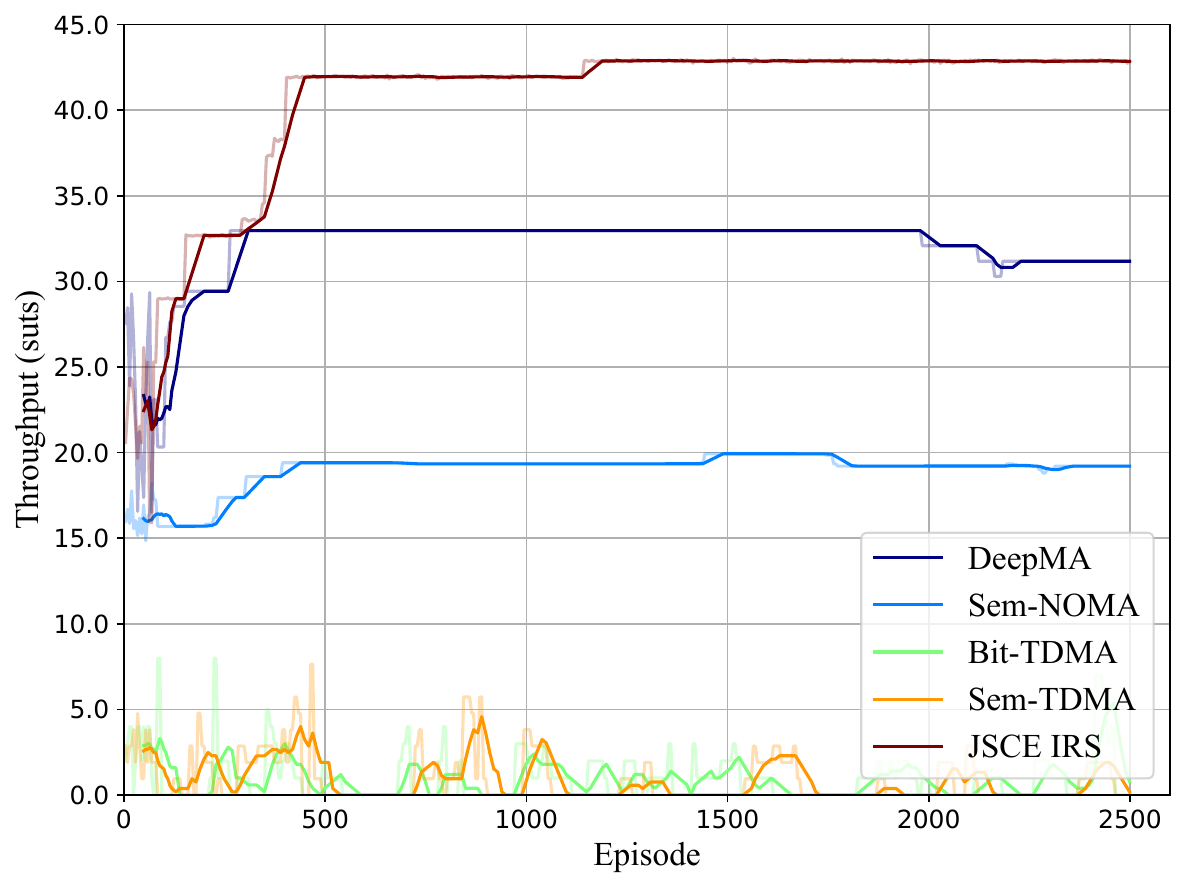}
    \caption{The training performance with different schemes. }
    \label{fig:reward}
    \vspace{-0.5cm}
\end{figure}

Figure~\ref{fig:reward} shows the training performance of the proposed schemes in multi-user semantic communication systems. The JSCE scheme achieves the best throughput performance. \textcolor{black}{In high-density deployment scenarios, strong resource coupling between users limits the transmission efficiency of the TDMA and NOMA schemes. However, the JSCE scheme leverages both semantic and spatial features to enhance the orthogonality between users, significantly improving the transmission performance. Moreover, JSCE outperforms DeepMA, validating the effectiveness of incorporating the IRS-controlled CSI as spatial features for semantic decoding. This allows JSCE to substantially reduce the size of the semantic model. For instance, in a 5-user scenario, DeepMA requires each user to maintain a 147.7 MB semantic module while JSCE reduces the model size to 27.29 MB per user, achieving an 80\% reduction.}
%This is because TDMA allows for only one transmission per time slot, and semantic-based NOMA primarily exploits the orthogonality of the channel. \textcolor{black}{In high-dense deployment and non-IRS scenarios, semantic-based NOMA must reduce the number of accessing users to ensure the quality of the transmission.} In summary, both TDMA and semantic-based NOMA have limited the simultaneous transmission capacity, necessitating a trade-off in system throughput to ensure user fairness. DeepMA allows simultaneous transmission for multiple pairs, outperforming semantic-based NOMA and TDMA. However, DeepMA requires paired encoder-decoders and lacks possible network topology that JSCE can provide. In a 5-user scenario, DeepMA's model size is 147.7 MB per user, while JSCE's is only 27.29 MB. JSCE's model has been reduced by 80\%, and this proportion will increase with more users.
\vspace{-0.5cm}
\section{Conclusion}
\vspace{-0.1cm}
In this paper, we have proposed a JSCE scheme for an IRS-assisted semantic communication system, enabling efficient simultaneous transmission for multiple users. We have introduced an XD-DRL framework to maximize the users' semantic throughput by jointly optimizing the users’ scheduling, IRS's passive beamforming, and semantic encoding strategies. The original problem is decomposed into two subproblems and solved by using backpropagation and DRL, respectively. \textcolor{black}{Numerical results have demonstrated that our proposed JSCE scheme enhances both semantic and spatial orthogonality, achieving greater semantic throughput compared to conventional benchmark schemes.}

% if have a single appendix:
%\appendix[Proof of the Zonklar Equations]
% or
%\appendix  % for no appendix heading
% do not use \section anymore after \appendix, only \section*
% is possibly needed

% use appendices with more than one appendix
% then use \section to start each appendix
% you must declare a \section before using any
% \subsection or using \label (\appendices by itself
% starts a section numbered zero.)
%

% Can use something like this to put references on a page
% by themselves when using endfloat and the captionsoff option.
\ifCLASSOPTIONcaptionsoff
  \newpage
\fi
\vspace{-0.1cm}

\bibliographystyle{IEEEtran}
% \bibliography{bibtex/bib/IEEEexample}
% Generated by IEEEtran.bst, version: 1.14 (2015/08/26)

% that's all folks
\end{document}